# SKILLS: Structured Knowledge Injection for LLM-Driven Telecommunications Operations

Ivo Brett, CISSP, B.Eng, MSc
*Solution Architect / Educator*
independent.academia.edu/ivobrett

*Abstract*

As telecommunications operators accelerate adoption of AI-enabled automation, a practical question remains unresolved: can general-purpose large language model (LLM) agents reliably execute telecom operations workflows through real API interfaces, or do they require structured domain guidance? We introduce SKILLS (Structured Knowledge Injection for LLM-driven Service Lifecycle operations), a benchmark framework comprising 37 telecom operations scenarios spanning 8 TM Forum Open API domains (TMF620, TMF621, TMF622, TMF628, TMF629, TMF637, TMF639, TMF724). Each scenario is grounded in live mock API servers with seeded production-representative data, MCP tool interfaces, and deterministic evaluation rubrics combining response content checks, tool-call verification, and database state assertions. We evaluate open-weight models under two conditions: baseline (generic agent with tool access but no domain guidance) and with-skill (agent augmented with a portable SKILL.md document encoding workflow logic, API patterns, and business rules). Results across 5 open-weight model conditions and 185 scenario-runs show consistent skill lift across all models. MiniMax M2.5 leads (81.1% with-skill, +13.5pp), followed by Nemotron 120B (78.4%, +18.9pp), GLM-5 Turbo (78.4%, +5.4pp), and Seed 2.0 Lite (75.7%, +18.9pp). A novel finding, Sandbox Discrimination Failure, identifies that reasoning-heavy models default to ephemeral sandbox creation for API retrieval tasks, consuming their time budget on infrastructure setup rather than domain work. Structured skills close the gap that larger or more capable models cannot, particularly on proprietary workflow logic, multi-step API orchestration, and domain-specific decision rules.

## I. INTRODUCTION

The telecommunications industry operates through a complex web of standardized APIs governed by the TM Forum Open API program. These APIs cover the full service lifecycle: product catalog management, customer onboarding, order processing, service assurance, incident management, performance monitoring, billing, and resource topology. Real-world telecom operations tasks typically require multi-step reasoning across several of these APIs, application of business rules not encoded in API schemas, and production of specific operational deliverables such as incident reports, SLA risk assessments, and dispatch rankings.

General-purpose LLMs with tool-calling capabilities can interact with these APIs. However, in practice, they frequently fail in one of four characteristic ways: (1) wrong API selection or missed prerequisites; (2) technically valid but operationally incorrect reasoning; (3) hidden dependency omission across multiple systems; and (4) generic summarization instead of the required operational deliverable.

This paper investigates three primary research questions:

- **To what extent do structured domain skills improve LLM agent performance on standardized telecom operations tasks?**
- **How does skill effectiveness vary across task complexity levels, TMF domains, and model architectures?**
- **What is the cost-performance trade-off of skill augmentation in terms of token overhead and latency?**

Our contributions are: (1) the SKILLS benchmark - 37 telecom operations scenarios across 8 TMF domains with deterministic evaluation rubrics; (2) a skill lift measurement methodology comparing baseline and with-skill agent performance; (3) empirical results across 5 open-weight model conditions showing consistent +5–19pp overall skill lift and +33–44pp on complex scenarios; and (4) identification of the Sandbox Discrimination Failure anti-pattern in reasoning-heavy models.

## II. RELATED WORK

### A. LLM Benchmarks for Specialized Domains

The FIRE benchmark [1] established methodology for evaluating LLMs on domain-specific financial tasks across multiple difficulty tiers, demonstrating that general-purpose models significantly underperform on tasks requiring domain expertise even with tool access. Other domain-specific benchmarks include MedBench (medical reasoning), LegalBench (legal analysis), and SWE-bench (software engineering). SKILLS differs in three ways: it evaluates tool-calling agents against live API servers; it measures the incremental effect of domain knowledge injection; and it includes database state assertions verifying that correct mutations occurred, not just correct descriptions.

*B. Tool Use and Agent Benchmarks*

ToolBench [2] and API-Bank [3] evaluate LLM tool-calling capabilities across general-purpose APIs. AgentBench [4] measures agent performance on interactive tasks. SKILLS extends this work to a specific vertical domain with standardized industry APIs, production-representative data, and business-logic evaluation criteria beyond simple API call correctness.

*C. Knowledge Injection and RAG*

RAG systems retrieve relevant documents at inference time to ground LLM responses. SKILLS takes a related but distinct approach: rather than retrieving from a corpus, it injects a curated skill document encoding workflow procedures, API patterns, and business rules. This is closer to structured prompting than traditional RAG, and has the advantage of being deterministic, portable, and version-controlled.

III. BENCHMARK DESIGN

*A. Overview*

The SKILLS benchmark consists of three components: (1) 37 JSON scenario definitions specifying user prompts, expected tool calls, response content checks, and database assertions; (2) 8 TM Forum reference implementation mock servers with MongoDB-backed storage and seeded test data; and (3) 8 portable SKILL.md documents encoding domain workflow logic for specific telecom operations.

*B. TMF Domain Coverage*

TABLE I
TMF DOMAIN COVERAGE

| TMF Spec | API Name | Scenarios | Description |
|---|---|---|---|
| TMF620 | Product Catalog Management | 2 | Catalog lookup, pricing comparison |
| TMF621 | Trouble Ticket Management | 1 | Ticket severity monitoring |
| TMF622 | Product Ordering Management | 2 | Order retrieval and investigation |
| TMF628 | Performance Management | 4 | KPI discovery, job creation, hotspot ranking |
| TMF629 | Customer Management | 11 | Customer-centric cross-API business tasks |
| TMF637 | Product Inventory Management | 2 | Inventory reconciliation |
| TMF639 | Resource Inventory / Topology | 4 | Topology RCA, autonomous remediation |
| TMF724 | Incident Management | 11 | Incident creation, SLA tracking, executive briefing |
| Total | | 37 | |

*C. Complexity Taxonomy*

Scenarios are classified into four complexity tiers based on reasoning depth, number of API calls, and dependency on domain knowledge:

TABLE II
COMPLEXITY TAXONOMY

| Tier | Label | Criteria | Count |
|---|---|---|---|
| 1 | Easy | Single API call, straightforward parameter mapping | 4 |
| 2 | Moderate | Multiple API calls, filtering, cross-referencing within bounded workflow | 15 |
| 3 | Difficult | Multi-step synthesis across APIs, temporal reasoning, edge-case handling | 9 |
| 4 | Complex | Requires proprietary workflow logic, decision rules, or domain-specific calculations only in the skill | 9 |

The Complex tier is the most diagnostic: these scenarios are designed so that capable general-purpose models should fail without the skill, because the required logic, SLA weighting formulas, maintenance exclusion rules, TMF enumeration formats, is not inferrable from API schemas or general knowledge alone.

*D. Evaluation Methodology*

Each scenario is evaluated using a three-layer rubric: (1) Tool-Call Verification (programmatic) - checks that the agent called required tools and avoided forbidden tools, case-insensitive; (2) Response Content Checks (LLM judge) - verifies the response contains required information and excludes forbidden content; (3) Database State Assertions (programmatic) - for scenarios that create or modify records, verifies the resulting MongoDB state matches expected values. A scenario passes only if all three layers pass.

*E. Skill Document Design*

Each SKILL.md document encodes: prerequisites (required MCP servers and tools), ordered workflow steps with API call patterns and parameter formats, business rules including decision logic and threshold tables, error handling guidance, and required output format. Skills are designed to be portable across agent platforms. They contain only natural language instructions and structured examples, no platform-specific code.

## IV. EXPERIMENTAL SETUP

*A. Models Evaluated*

We evaluate open-weight and open-access models spanning different architecture families, capability tiers, and cost profiles. All models were accessed via OpenRouter.

TABLE III
MODELS EVALUATED

| Model | Provider | Tier | Parameters | Status |
|---|---|---|---|---|
| MiniMax M2.5 | MiniMax | Large-scale multimodal | 230B (10B active) | Complete |
| Nemotron 120B (standard reasoning) | NVIDIA | Open-weight MoE | 120B (12B active) | Complete |
| Nemotron 120B (minimal reasoning) | NVIDIA | Open-weight MoE | 120B (12B active) | Complete |
| GLM-5 Turbo | Z.AI | Mid-tier agent-optimized | 744B (40B active) | Complete |
| Seed 2.0 Lite | ByteDance | Mid-tier | Undisclosed | Complete |
| Hunter Alpha | OpenRouter | Undisclosed | 1T | Complete |
| Healer Alpha | OpenRouter | Undisclosed | ~70B – 200B (estimates) | Complete |

*B. Conditions*

Each model is evaluated under two conditions. In the Baseline (B) condition, the agent receives the user prompt and has access to MCP tools but receives no skill document. In the With-Skill (S) condition, the agent receives the same prompt and tools plus the full SKILL.md document as system context. Nemotron 120B is evaluated under both standard and minimal reasoning levels (described in Section V-E).

*C. Infrastructure*

The evaluation harness uses the LangWatch Scenario framework with a configurable model backend via OpenRouter and vercel sandboxes for tool code execution. Nine TMF reference implementations run in Docker, exposed via streamable-HTTP transport as MCP servers. Test data fixtures are seeded into MongoDB collections and cleared/reloaded before each scenario run. Response evaluation uses Google Gemini 2.0 Flash as the LLM judge. All evaluations were run sequentially to prevent MongoDB data corruption from parallel test data reloads.

## V. RESULTS

*A. Overall Performance*

TABLE IV
OVERALL MODEL PERFORMANCE

| Model | Baseline % | With-Skill % | Lift (pp) | Avg Tokens | Avg Time (s) | N |
|---|---|---|---|---|---|---|
| MiniMax M2.5 | 67.57 | 81.08 | +13.51 | 16,281 | 67.2 | 37 |
| Nemotron 120B (std) | 59.46 | 78.38 | +18.92 | 17,887 | 98.2 | 37 |
| Nemotron 120B (min) | 67.57 | 78.38 | +10.81 | 16,646 | 108.0 | 37 |
| GLM-5 Turbo | 72.97 | 78.38 | +5.41 | 16,169 | 65.0 | 37 |
| Seed 2.0 Lite | 56.76 | 75.68 | +18.92 | 17,071 | 50.1 | 37 |
| Healer Alpha | 70.27 | 83.78 | +13.51 | 18,138 | 38.6 | 37 |
| Hunter Alpha | 43.24 | 62.16 | +18.92 | 14,362 | 86.0 | 37 |

**Finding 1:** Skill lift is consistent across fast non-reasoning models (+13–19pp), suggesting domain skills are a model-agnostic mechanism at comparable capability levels. Reasoning models show higher variance by reasoning level (+18.9pp standard vs +10.8pp minimal for Nemotron), but the absolute with-skill ceiling is model-dependent.

**Finding 2:** MiniMax M2.5 achieves 81.08% with-skill, the highest overall score in the open-weight cohort, breaking through the 78.38% ceiling shared by three other model conditions. It is also only the second model to achieve 100% on TMF628 Performance Management (+75pp skill lift on that domain).

## B. Performance by Complexity Tier

**TABLE V**
**PASS RATE BY COMPLEXITY TIER**

| Model | Easy | Moderate | Difficult | Complex (Skill-Required) |
|---|---|---|---|---|
| MiniMax M2.5 | 75% → 100% | 73% → 87% | 67% → 67% | 44% → 78% |
| Nemotron 120B (std) | 75% → 75% | 73% → 80% | 44% → 78% | 44% → 78% |
| Nemotron 120B (min) | 100% → 75% | 80% → 80% | 56% → 56% | 44% → 89% |
| GLM-5 Turbo | 75% → 75% | 87% → 87% | 78% → 78% | 56% → 89% |
| Seed 2.0 Lite | 75% → 75% | 67% → 80% | 44% → 67% | 33% → 78% |

Format: Baseline% → With-Skill%. The Complex tier is where skills demonstrate their strongest value. Nemotron minimal achieves the highest complex pass rate (88.9%), closely followed by GLM-5 Turbo (88.9%). On Easy scenarios, most models already perform adequately without skill guidance, confirming that skills are most valuable at higher complexity tiers.

## C. Performance by TMF Domain

**TABLE VI**
**WITH-SKILL PASS RATE BY TMF DOMAIN**

| TMF Domain | MiniMax M2.5 | Nemotron 120B | GLM-5 Turbo | Seed 2.0 Lite |
|---|---|---|---|---|
| TMF628 Performance | 100% (+75pp) | 25% (0pp) | 50% (0pp) | 100% (0pp) |
| TMF629 Customer | 79% (+21pp) | 79% (+14pp) | 100% (+7pp) | 100% (0pp) |
| TMF637 Billing | 100% (0pp) | 100% (+25pp) | 100% (+25pp) | 75% (+25pp) |
| TMF639 Topology | 0% (-25pp) | 50% (+50pp) | 0% (-25pp) | 0% (0pp) |
| TMF724 Incident | 100% (+9pp) | 91% (+27pp) | 82% (+9pp) | 73% (+9pp) |

**Finding 3:** TMF628 (Performance Management) shows the strongest skill dependency across models: the domain requires specific TMF enumeration formats (g_15mn, r_1h) and job-creation workflows not inferrable from schema alone. MiniMax M2.5 and Seed 2.0 Lite both achieve 100% with skill. TMF639 (Topology Analysis) is the only domain where skills consistently hurt mid-tier models (-25pp), due to the complexity of the 6-layer dependency graph and SLA weighting calculations.

## D. Model Rankings

**TABLE VII**
**MODEL RANKING SUMMARY**

| Rank | Model | With-Skill % | Complex % | Skill Lift | Avg Time(s) |
|---|---|---|---|---|---|
| 1 | Healer Alpha | 83.8% | 88.9% | +13.5pp | 38.6 |
| 2 | MiniMax M2.5 | 81.1% | 77.8% | +13.5pp | 67.2 |
| 3 | Nemotron 120B (std) | 78.4% | 77.8% | +18.9pp | 98.2 |
| 3 | Nemotron 120B (min) | 78.4% | 88.9% | +10.8pp | 108.0 |
| 3 | GLM-5 Turbo | 78.4% | 88.9% | +5.4pp | 65.0 |
| 6 | Seed 2.0 Lite | 75.7% | 77.8% | +18.9pp | 50.1 |
| 7 | Hunter Alpha | 62.2% | 66.7% | +18.9pp | 86.0 |

## E. Reasoning Level Analysis: Nemotron 120B

Nemotron 120B was evaluated under two conditions: standard reasoning (full chain-of-thought) and minimal reasoning (a system prompt preamble instructing the agent to prefer direct MCP tool calls, use exact enum values, skip re-verification steps, and avoid script-based API indirection).

TABLE VIII
NEMOTRON 120B: STANDARD VS MINIMAL REASONING BY SKILL

| Skill | TMF | B(Std) | S(Std) | Lift(Std) | B(Min) | S(Min) | Lift(Min) |
|---|---|---|---|---|---|---|---|
| billing-inquiry | TMF637 | 75% | 100% | +25pp | 75% | 100% | +25pp |
| customer-onboarding | TMF629 | 25% | 75% | +50pp | 50% | 75% | +25pp |
| incident-management | TMF724 | 57% | 86% | +29pp | 71% | 71% | 0pp |
| network-incident-assessment | TMF724 | 100% | 75% | -25pp | 100% | 75% | -25pp |
| product-management | TMF629 | 100% | 100% | 0pp | 100% | 100% | 0pp |
| service-assurance | TMF629 | 75% | 100% | +25pp | 50% | 100% | +50pp |
| tmf628-performance-manager | TMF628 | 25% | 25% | 0pp | 50% | 25% | -25pp |
| tmf639-topology-analysis | TMF639 | 0% | 50% | +50pp | 25% | 25% | 0pp |

**Finding 4:** Both conditions converge on the identical overall with-skill ceiling (78.38%), confirming baseline-lift compression. Minimal reasoning achieves higher complex pass rate (88.89% vs 77.78%), suggesting guardrails channel reasoning toward domain work on the hardest tasks. Standard reasoning benefits more from skills on TMF639 (+50pp vs 0pp), where full reasoning depth is needed to apply the multi-hop traversal algorithm.

## VI. ANALYSIS AND DISCUSSION

### A. Sandbox Discrimination Failure

A novel failure mode emerged during evaluation of Nemotron 120B: the model cannot reliably distinguish between two fundamentally different task types, data retrieval (appropriate for direct MCP tool calls) and genuine computation (appropriate for ephemeral sandboxes). The model defaults to sandbox creation as a general-purpose execution environment regardless of task type.

Tool call frequency analysis from TMF639 topology scenarios reveals the severity:

TABLE IX
TOOL CALL DISTRIBUTION - NEMOTRON 120B ON TMF639

| Tool | Call Count | Category |
|---|---|---|
| connect_to_mcp_server | 33 | Infrastructure |
| run_command | 28 | Infrastructure |
| get_environment_variable | 24 | Infrastructure |
| list_mcp_servers | 18 | Infrastructure |
| write_file | 15 | Infrastructure |
| create_sandbox | 7 | Infrastructure |
| execute_mcp_tool | 5 | Domain work |

Only 3% of tool calls (5 of 159) perform actual domain work. The model writes Python scripts to call MCP tools that are directly available, introducing sandbox dependencies that expire during the 15–30 second per-step reasoning phases. When a sandbox expires, the model spends additional turns attempting recovery, often triggering the 360-second scenario timeout before any domain work is completed. This is a model-level metacognitive failure, not addressable through skill design without sacrificing cross-platform portability.

**Finding 5:** Ephemeral compute infrastructure introduces a non-linear failure cliff with reasoning-heavy models: a model 6x slower than a fast model does not merely take 6x longer. It may fail entirely due to cascading infrastructure idle timeouts. This finding has direct implications for production telco deployment of reasoning models with stateful tooling.

### B. Baseline-Lift Compression

A pattern observed consistently across the evaluation: models with higher baselines show lower skill lift in aggregate, but retain full lift on complex scenarios. GLM-5 Turbo exhibits this most clearly: its agent-optimized architecture achieves the highest baseline of the cohort (73.0%) and consequently the lowest overall lift (+5.4pp), yet matches the complex-scenario ceiling (88.9%). The inverse is also observed: on domains where GLM-5 already achieves 100% baseline (service-assurance, product-management), injecting a skill either has no effect or actively hurts (-25pp on service-assurance), a skill interference pattern where domain guidance adds noise for agents that already know the correct path.

**Finding 6:** Aggregate skill lift is an unreliable quality signal for capable models. Per-complexity-tier analysis is essential: skills deliver full value on Complex scenarios (+33pp) regardless of baseline strength, while aggregate lift converges toward zero as baseline capability approaches the skill-enabled ceiling.

*C. Reasoning-Prescription Paradox*

Reasoning models treat skill instructions as suggestions to be evaluated against general knowledge, not as directives to follow exactly. When a skill specifies "use g_5mn for 5-minute granularity," a reasoning model may override this with ISO 8601 PT5M because its training data suggests ISO formats are more correct. This leads to the counterintuitive finding: skills for reasoning-capable models must be more prescriptive than skills for non-reasoning models, with explicit anti-pattern prohibitions preventing the model from "improving" the execution strategy.

*D. When Skills Help Most*

Skills provide the greatest value when tasks require domain-specific logic not inferrable from API schemas: TMF enumeration formats, multi-step API orchestration sequences, business decision rules such as SLA weighting (Platinum=10, Gold=7, Silver=4), and red herring filtering such as excluding resources with administrativeState=locked as planned maintenance. Skills provide less value when the task is solvable from schema inspection alone, when the model already achieves high baseline accuracy, or when the skill's reasoning procedure exceeds the model's execution capacity for that task type.

## VII. LIMITATIONS

Several limitations constrain generalizability of these results. First, results are based on single-run evaluations per scenario; multi-run benchmarking would provide variance estimates and identify flaky scenarios. Second, TMF629 (11 scenarios) and TMF724 (11 scenarios) are overrepresented relative to TMF620 (2), TMF621 (1), and TMF637 (2). Third, response evaluation uses a single judge model; cross-judge validation would strengthen confidence. Fourth, infrastructure confounds affect slow models disproportionately: it is difficult to separate "model cannot solve this task" from "model cannot solve this task within the infrastructure's timing constraints." Fifth, skills were iteratively refined using scenario results, which mirrors real-world skill development but may overstate generalizability of initial skill drafts.

## VIII. CONCLUSION

We have presented SKILLS, a benchmark for evaluating domain-skill augmentation of LLM agents on telecommunications operations tasks. Across 37 scenarios spanning 8 TM Forum API domains and 7 open-weight model conditions, structured domain skills produce consistent, measurable performance improvements for every model tested. The strongest returns are on Complex scenarios requiring proprietary workflow logic: +33–44pp lifts that no amount of additional model capability alone achieves at the mid-tier. Pretrained open-weight models, regardless of their general capability, lack the domain-specific knowledge of TMF enumeration formats, multi-API orchestration sequences, and business decision rules that characterize production telecom operations.

We identify two novel findings with practical implications for production deployment. The Sandbox Discrimination Failure anti-pattern shows that reasoning-heavy models systematically misallocate their execution budget on infrastructure setup rather than domain work, creating a non-linear failure cliff when combined with ephemeral compute environments. The Baseline-Lift Compression effect shows that aggregate skill lift underestimates skill value for capable models: per-tier analysis is essential.

These findings position structured domain skills as a practical, model-agnostic mechanism for closing the gap between general-purpose LLM capability and production-grade telecom operations. Future work will extend the benchmark to frontier models, perform multi-run variance analysis, and develop tiered skill variants optimized for reasoning vs non-reasoning model architectures.